\documentclass[epj]{svjour}
\usepackage{graphics}
\begin{document}
\title{Effective Heisenberg exchange integrals of diluted magnetic semiconductors determined within realistic multi-band tight-binding models}
\titlerunning{Effective Heisenberg Exchange Integrals}
\author{Stefan Barthel\inst{1} \and Gerd Czycholl\inst{1} \and Georges Bouzerar\inst{2,3}
}                     
%
%
\institute{Institute for Theoretical Physics, University of Bremen, Otto-Hahn-Allee 1, D-28359 Bremen, Germany \and Institut N\'eel, 25 avenue des Martyrs, B.P. 166, 38042 Grenoble Cedex 09, France \and School of Engineering and Science, Jacobs University Bremen, Campus Ring 1, D-28759 Bremen, Germany}
\date{Received: date / Revised version: date}
%
\abstract{
Diluted magnetic semiconductors (DMS) like Ga$_{1-x}$Mn$_{x}$As are described by a realistic tight-binding model (TBM) for the (valence) bands of GaAs, by a Zener (J-)term modeling the coupling of the localized Mn-spins to the spins of the valence band electrons, and by an additional potential scattering (V-) term due to the Mn-impurities.  We calculate the effective (Heisenberg) exchange interaction between two Mn-moments mediated by the valence electrons. The influence of the number of bands taken into account (6-band or 8-band TBM) and of the potential (impurity) scattering V-term is investigated. We find that Ð for realistic values of the parameters Ð the indirect exchange integrals show a  long-range, oscillating  (RKKY-like) behavior, if the V-term is neglected, probably leading to spin-glass behavior rather than magnetic order. But by including a V-term of a realistic magnitude the exchange couplings become short ranged and mainly positive  allowing for the possibility of ferromagnetic order. Our results are in good agreement with available results of ab-initio treatments.
} 
\maketitle
\section{Introduction}
\label{intro}
Dilute magnetic semiconductors (DMS) have risen a lot of interest during the last decade because of their possible use for spintronic devices \cite{RevModPhys.76.323} (e.g.  as spin field effect transistor or for spin-polarized light-emitting diodes). A key goal is the achievement of Curie temperatures $T_C$ well above room temperature; the largest $T_C$ obtained so far is, however, only of the magnitude of about 173 K and has been obtained for Ga$_{1-x}$Mn$_x$As \cite{PhysRevB.72.165204}. The possibility of a much larger $T_C$ has been predicted theoretically, and even a $T_C  \approx 700$ K has been predicted for GaMnN \cite{Dietl11022000}, but this could not yet be confirmed experimentally. Furthermore, it has been demonstrated that the theoretical prediction of such a  huge $T_C$ is the consequence of several drastic approximations, namely (1) perturbative treatment, (2) mean-field approximation and (3) virtual crystal approximation for the treatment of disorder \cite{PhysRevLett.93.137202,0295-5075-69-5-812}. Therefore, improved and more reliable theories of DMS are necessary, and so far two different kinds of approaches have been applied, namely (1) realistic band structure model studies (kp-, Kohn-Luttinger, empirical tight-binding models) \cite{RevModPhys.78.809} and (2) material specific ab-initio calculations \cite{RevModPhys.82.1633}. Recently it has been shown that the essential properties of DMS can already be described  within simplified model studies \cite{0295-5075-92-4-47006,1367-2630-13-2-023002} using a one-band model. Therefore, a model with a more realistic band structure and a non-perturbative treatment of the exchange coupling between the carriers and the magnetic moments should be able to realistically capture already the essential physical properties of DMS.

In this paper we start from a realistic empirical tight-binding model (ETBM) for the electronic properties of GaAs and take into account the Mn-impurities by two additional terms, namely the local exchange coupling $J_{pd}$ between the spins of the (valence p-) electrons (or the holes) and the localized magnetic Mn-(d-shell) moments (ÔÕJ-termÕÕ) and the impurity potential (ÔÕV-termÕÕ) provided by the Mn-ions. A similar model was also used in recent Monte-Carlo (MC) simulations \cite{PhysRevLett.99.057207}. However, in this previous work only the three valence bands (per spin direction)  of GaAs were taken into account, furthermore the V-term was completely neglected and the whole investigation was restricted to relatively small systems.   In the present paper we also start from a realistic ETBM description of the electronic properties of GaAs. The TB-parameters are determined so that the most important known band structure features (energy gap, effective masses, Luttinger parameters, etc.) are well reproduced. We consider both, a six-band model (i.e. only the three valence bands per spin direction as investigated in \cite{PhysRevLett.99.057207}) and a more realistic eight-band model (including also one conduction band). We take into account a finite concentration $x$ of Mn-ions, which provide for a local impurity potential $V$, a local exchange coupling $J_{pd}$ and simultaneously serve as dopants (acceptors), and treat the disorder exactly. Here the kernel-polynomial method (KPM) \cite{RevModPhys.78.275} is employed, which allows for the essentially exact calculation of the density of states of the disordered system for relatively large (finite) systems. The KPM can also be used for the calculation of the effective (long-ranged, Heisenberg) exchange interaction $J_{ij}$ between two Mn-ions at sites $i$ and $j$ mediated by the free carriers (i.e. the electrons or rather the holes in the valence bands), starting from the Lichtenstein-Katsnelson formula \cite{Liechtenstein198765}.

 For realistic choices of the parameters ($J_{pd}S= 3$ eV, Mn-concentration $x= 5 \%$)  we obtain a RKKY-like, long-ranged oscillating behavior of $J_{ij}$, if the potential scattering $V$-term is neglected, for both, the 6-band and the 8-band model. In particular, $J_{ij}$ may be positive or negative (ferromagnetic or antiferromagnetic) depending on the distance between two Mn-ions and on the configuration. Therefore, no magnetic order can be expected but rather a spin-glass behavior. On the other hand, if the V-term is properly included (of the correct magnitude so that the bound state due to Mn impurities in the gap is reproduced) we obtain a more short ranged $J_{ij}$, which is mainly positive so that ferromagnetic order should become possible. Altogether the 8-band description is, of course, more reliable and realistic and leads, in particular, to valence bands of a realistic band width in the simplest ETBM description. We also demonstrate that for the 8-band model and a finite $V$ our result for $J_{ij}$ is already in rather good agreement with the corresponding results of ab-initio calculations.

\section{Theory}
\label{sec:1}
Our approach to theoretically describe the physics of the DMS material Ga$_{1-x}$Mn$_x$As combines an empirical tight-binding model (ETBM) for the host semiconductor with additional terms describing the nonmagnetic impurity scattering and a spin-spin interaction term, which couples impurity spins to carrier spins. This kind of model is referred to as the V-J model in the literature, see Refs.\cite{0295-5075-78-6-67003,springerlink:10.1140/epjb/e2011-20320-x}, but we want to go beyond a single-band description. Neglecting the potential scattering $V$ our ansatz becomes equivalent to the Zener model. From the corresponding Green's functions one is able to calculate the effective Mn-Mn exchange integrals in the presence of disorder according to the Lichtenstein-Katsnelson formula, Ref.\cite{Liechtenstein198765}. We treat the disorder exactly (for finite systems) and perform the calculations for different realizations of the disordered system and, therefore, obtain a distribution of and calculate the average of the couplings $J_{ij}$. These couplings can then be used as the exchange interaction of a disordered Heisenberg model to calculate magnetic properties of the system. After this short overview about the methodology applied, we are going to be more specific.

\subsection{Multiband V-J model}
\label{sec:1a}
We study the following hamiltonian:
\begin{equation}
\label{eq:h_tb}
\hat{H} = \sum_{ij,\alpha\beta,\sigma}t_{ij}^{\alpha\beta}
\hat{c}_{i\alpha\sigma}^{\dagger}\hat{c}_{j\beta\sigma}+\sum_{i,\alpha,\sigma}p_iV_{\alpha}\hat{n}_{i\alpha\sigma}
+\sum_{i,\alpha}p_iJ_{\alpha}\hat{\bf{S}}_i\hat{\bf{s}}_{i,\alpha}.
\end{equation}
Here, the first term corresponds to the ETBM description of the relevant energy bands of GaAs with $t_{ij}^{\alpha\beta}$ being the hopping integral of an electron from lattice site $j$ to $i$ in the orbital $\beta$ to $\alpha$. The corresponding creation (annihilation) operator is $\hat{c}_{i\alpha\sigma}^{\dagger}$ $(\hat{c}_{j\beta\sigma})$ with the additional spin-index $\sigma\in\{\uparrow,\downarrow\}$. At this point it is convenient to discuss the tight-binding parametrization for the $t_{ij}^{\alpha\beta}$ and the basis in more detail. One possibility is to set up the hamiltonian for the full atomistic lattice and use $s,p,d$-orbitals because the chemical bond of GaAs is $sp^3$-hybridized and the Mn$^{2+}$ impurities under consideration introduce $d$-orbitals into the problem. These kinds of approaches were already studied in the work of Tang and Flatt\'e \cite{PhysRevLett.92.047201} without $d$-orbitals, by Ma\ifmmode \check{s}\else \v{s}\fi{}ek \textit{et al.} \cite{PhysRevB.75.045202} including the $d$-orbitals as well compared in Ref.\cite{PhysRevB.78.085211} by Turek \textit{et al.}. We use an alternative, simpler version of an ETBM starting from well localized (Wannier) orbitals on the fcc lattice and can use two different analytical parametrizations for the $t_{ij}^{\alpha\beta}$. The first one uses only $p$-orbitals describing the valence bands of GaAs and was already studied in Ref.\cite{PhysRevLett.99.057207}, while the second one uses the $sp^3$-basis as first suggested by Loehr in Ref.\cite{PhysRevB.50.5429} and was also applied to ferromagnetic semiconductor superlattices \cite{PhysRevB.64.245207}. The ETBMs which operate on a Bravais lattice are usually referred to as effective-bond orbital model (EBOM) and have the benefit, that they do analytically include the $\mathbf{k\cdot p}$-hamiltonian in the limit $k\to0$. This means, in particular, that the $t_{ij}^{\alpha\beta}$ are given by analytic expressions in terms of (anisotropic) effective masses $m_{c}$, Luttinger parameters $\gamma_{i}$, bandgaps $E_g$, critical point energies, the Kane parameter $E_p$, etc. We use the input parameters given in Tab.(\ref{tab:1}) for the 6-band and 8-band ETBM, but neglect the spin-orbit splitting. There is no a-priori justification for doing this, since for GaAs $\Delta_{\mathrm{so}}=341$ meV is relatively large. Nevertheless we want to compare our results for the effective exchange integrals $J_{ij}$ to results from ab-initio tight-binding linear muffin-tin orbital theory (TB-LMTO) with the coherent potential approximation (CPA) which also neglect spin-orbit interaction, see Ref.\cite{PhysRevB.69.115208} for details.

\begin{table}
\caption{Model parameters for GaAs used in the calculations. Unchecked quantities (-) are not included in the respective analytical tight-binding parametrization.}
\label{tab:1}
\centering
\begin{tabular}{lr|crcc}
\hline\noalign{\smallskip}
 Parameter & & Value & & 6-bands & 8-bands \\
\noalign{\smallskip}\hline\noalign{\smallskip}
$a$ & \AA & 5.64 & \cite{PhysRevLett.99.057207} & $\surd$ & $\surd$ \\
$\Delta_{so}$ & eV & 0.341& \cite{vurgaftman:5815} & - & - \\
$m_{c}$ & $m_{0}$ & 0.067 & \cite{vurgaftman:5815} & - & $\surd$ \\
$\Gamma_{1c}$ - $\Gamma_{15v}$ & eV & 1.519 & \cite{vurgaftman:5815} & - & $\surd$ \\
$X_{1c}$ & eV & 2.1 & & - & $\surd$ \\
$X_{5v}$ & eV & -3 & & - & $\surd$ \\
$X_{3v}$ & eV & -7 & & - & $\surd$ \\
$\gamma_{1}$ & & 6.85 & \cite{PhysRevLett.99.057207} & $\surd$ & $\surd$ \\
$\gamma_{2}$ & & 2.1 & \cite{PhysRevLett.99.057207} & $\surd$ & $\surd$ \\
$\gamma_{3}$ & & 2.9 & \cite{PhysRevLett.99.057207} & $\surd$ & $\surd$ \\

\noalign{\smallskip}\hline
\end{tabular}
\end{table}


The second term in Eq.(\ref{eq:h_tb}) describes the nonmagnetic impurity scattering with the short-range on-site potential $V_{\alpha}$ and $\hat{n}_{i\alpha\sigma}$ is the occupation operator. For the simulation of the disorder the variable $p_i$ takes the value 1 with probability $x$ and 0 with probability $1-x$ where $x$ denotes the impurity concentration of the sample for each lattice site $i$. This contribution to the hamiltonian does also arise from the Schrieffer-Wolf transformation \cite{PhysRev.149.491} of the Anderson hamiltonian, Ref.\cite{PhysRev.124.41}.
We use $V_{\alpha}=V$ for all orbitals from now on in order to keep the model parameters at bare minimum and its magnitude will be physically motivated and discussed in the results. Of course this part of the hamiltonian could be augmented by a nearest-neighbor Coulomb potential similar as done in Ref.\cite{PhysRevLett.105.227202}, but we choose to study the on-site case solely because it prevents introducing even more parameters in our study.

The last term of Eq.(\ref{eq:h_tb}) couples impurity spins $\hat{\mathbf{S}}_i$ to carrier spins $\hat{\mathbf{s}}_{i,\alpha}$ of the host system via a local contact-interaction with coupling constant $J_{\alpha}$ and we have to deal with quantum mechanical spin $5/2$ operators for the Mn$^{2+}$ impurities and $1/2$ for the carriers (holes). In the case of Ga$_{1-x}$Mn$_x$As the fundamental magnetic coupling mechanism is the coupling $J_{pd}$ between the spins of the (p-like) valence electrons (holes) and the impurity spins (formed by the Mn-d-shells); of course, there is also a coupling $J_{sd}$ between the (s-like) conduction electron spins and the Mn-spins. The magnitude is about $J_{pd}\approx1.2$ eV and $J_{sd}\approx0.02$ eV, see Refs.\cite{PhysRevLett.95.017204,PhysRevB.72.235313}, and we decided to use $J_{\alpha}=J_{pd}$ for all orbitals which is again due to our constraint to keep the model paramaters at minimum.
Now we must explain how we treat this spin-spin interaction term in our approach. Our calculations are performed for $T=0$ K and thus transverse impurity spin-fluctuations should be absent. We must then only consider the $z$-component of the spin-operators and replace $\hat{S}_{i}^{z}\rightarrow\langle\hat{S}_{i}^{z}\rangle=S$ by its expectation value for the calculations of the electronic properties only.
The spin-spin interaction term is then given by ($\hbar=1$):
\begin{equation}
\frac{J_{pd}S}{2}\sum_{i,\alpha}p_i(\hat{\bf{n}}_{i,\alpha,\uparrow}-\hat{\bf{n}}_{i,\alpha,\downarrow}).
\end{equation}
Finally we are left with a single-particle hamiltonian describing non-interacting electrons on a lattice via kinetic energy, magnetic and nonmagnetic scattering which we have to solve for each random configuration of substitutional disorder.

\begin{figure}
\centering
\resizebox{0.4\textwidth}{!}{\includegraphics{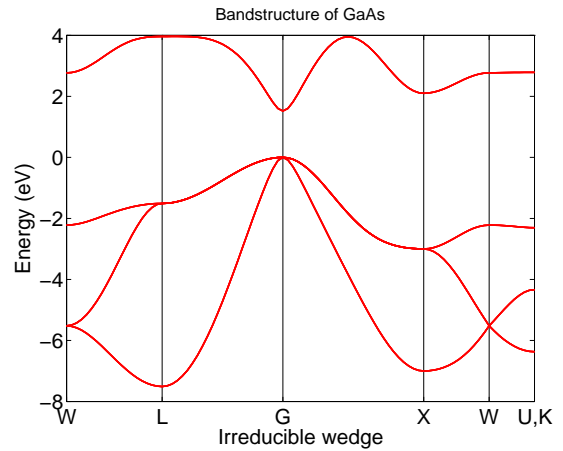}}
\caption{Bandstructure of bulk GaAs calculated within the 8-band model without spin-orbit splitting.}
\label{fig:1}
\end{figure}

In order to get an impression of the quality of the parametrization for bulk GaAs, we have plotted the bandstructure along the corresponding irreducible wedge for the 8-band model in Fig.(\ref{fig:1}). Due to the analytical expression for the hopping matrix elements, the experimental bandgap and effective masses at the $\Gamma$-point are reproduced, see Ref.\cite{vurgaftman:5815} for an overview, as well as the critical point energies at the $X$-point can be adjusted. In comparison to first-principle bandstructures in e.g. Refs.\cite{PhysRevB.52.4896,shimazaki:224105} the $\approx$12 eV bandwidth of the 8-band model is in reasonable agreement and the position of the $X$-point energies lies within the values reported in the literature, see e.g. Ref.\cite{PhysRevB.52.4896} for a comparison. Deviations in terms of energetic position and curvature start to appear for larger $k$-values along the $W-L$ path and along $X-W-K$ as expected. In the interval $L-\Gamma-X$ the agreement is better, despite the first-principles calculation in Ref.\cite{PhysRevB.52.4896} has difficulties to reproduce the experimental bandgap. The bandwidth of the 6-band model is approximately $\approx$30 eV and the agreement is only reasonable in the vicinity of the $\Gamma$-point, since the $X$-point energies are not pinned.

\subsection{Limitations of the model}
The TB-parameters used in our calculations are chosen so that the pure GaAs band structure is reasonably reproduced and that the Mn-impurities are properly described. For the latter purpose the impurity  parameters are chosen so that the bound Mn impurity state in the band gap of GaAs is reproduced, and these parameters (chosen for the single Mn impurity within GaAs) are used also for finite concentrations $x$ of Mn. One should be aware of the fact that this treatment is justified only in the low concentration limit of small $x$. Though in principle the method can be applied to arbitrary $x$ (even to the case $x=1$ of pure MnAs, see e.g. Ref.\cite{PhysRevB.62.15553}), the parameters are not appropriate for too large $x$. One has, for instance,  to choose and fit also off-diagonal hopping matrix elements appropriate for MnAs if interested in higher Mn-concentrations $x$

\subsection{Evaluation of exchange integrals}
\label{sec:1b}
Having solved the electronic problem, one is able to calculate the effective Mn-Mn exchange interaction between two impurities
located at different lattice sites $\mathbf{r}_i$ and $\mathbf{r}_j$ according to the expressions derived in Ref.\cite{Liechtenstein198765}:
\begin{equation}
\label{eq:Jij_full}
J_{ij}=\frac{1}{4\pi S^2}\Im\int\limits_{-\infty}^{\infty}f(\omega)\mathrm{Tr}_{\alpha}
\{\underline{\underline{\hat{\Sigma}}}_{i}(\omega)\underline{\underline{\hat{G}}}_{ij}^{\uparrow}(\omega)
\underline{\underline{\hat{\Sigma}}}_{j}(\omega)\underline{\underline{\hat{G}}}_{ji}^{\downarrow}(\omega)\}
\mathrm{d}\omega.
\end{equation}
Thereby vertex corrections have been neglected, which is justified according to Ref.\cite{PhysRevLett.76.4254}.
Here, $f(\omega)$ is the Fermi-function, $\mathrm{Tr}_{\alpha}$ denotes the trace over orbital indices and $\underline{\underline{\hat{G}}}_{ij}^{\sigma}(\omega)$ is the one-particle Green's function matrix with respect to the orbital basis in the spin-sector $\sigma$. According to the approximations outlined earlier, the self-energy reduces to $\underline{\underline{\hat{\Sigma}}}_{i}$ = J$_{pd}$S$\cdot\underline{\underline{\mathrm{\hat{1}}}}$, so that Eq.(\ref{eq:Jij_full}) can be written in a form where only products of two Green's functions have to be evaluated:
\begin{equation}
\label{eq:Jij_approx}
J_{ij}=\sum_{\alpha\beta}\underbrace{\frac{(J_{pd})^{2}}{4\pi}\Im\int\limits_{-\infty}^{\infty}
f(\omega)\hat{G}_{ij}^{\uparrow,\alpha\beta}(\omega)\hat{G}_{ji}^{\downarrow,\beta\alpha}(\omega)\mathrm{d}\omega}_{:=J_{ij}^{\alpha\beta}}
\end{equation}
Via the Fermi function $J_{ij}$ depends on the position of the Fermi energy, which has to be determined for the given number of carriers, which again depends on the number of holes $n_{h}$ introduced by the Mn$^{2+}$-doping. Though, in principle, the hole concentration could be less than the Mn-concentration $x$ (because of anti-site effects, i.e. Mn-ions occupying As-sites), we will assume $n_{h}=1$ for each Mn-impurity, because otherwise also additional anti-site impurity scattering terms would have to be included.

The $J_{ij}$ according to Eq.(\ref{eq:Jij_approx}) which depend on the actual configuration of the disordered system and are, therefore, subject to a probability distribution function, can be used as input parameters for a disordered Heisenberg model: 
\begin{equation}
\label{eq:heisenberg}
\hat{H}_{Heis}=-\sum_{i\neq j}p_{i}p_{j}J_{ij}\hat{\mathbf{S}}_{i}\hat{\mathbf{S}}_{j}.
\end{equation}

\subsection{Kernel polynomial method}
\label{sec:1c}
We want to study the electronic problem rigorously (for a finite system), thereby treating the disorder (impurity scattering) exactly. Since the exact diagonalization scales with the third power of the dimension of the Hamiltonian, we employ the kernel polynomial method (KPM), which essentially scales linear with the dimension of the Hamiltonian. The KPM, see the review by Weisse \textit{et al.} \cite{RevModPhys.78.275}, can be used to calculate the total density of states of the disordered system. It can also be used to calculate traces over products of matrix elements of operators. Therefore, also the Heisenberg exchange interaction in Eq.(\ref{eq:heisenberg}) can be calculated according to:
\begin{eqnarray}
\label{eq:Jij_kpm}
J_{ij}^{\alpha\beta}=\frac{(J_{pd})^{2}}{4}\int\limits_{-\infty}^{\infty}\mathrm{d}\omega\int\limits_{-\infty}^{\infty}\mathrm{d}\omega^{\prime}\frac{f(\omega^{\prime})-f(\omega)}{\omega-\omega^{\prime}}\cdot\nonumber\\
\Re\{\langle j\beta\uparrow\vert\hat{\delta}(\omega-\hat{H})\vert i\alpha\uparrow\rangle\cdot\langle i\alpha\downarrow\vert\hat{\delta}(\omega^{\prime}-\hat{H})\vert j\beta\downarrow\rangle\}.
\end{eqnarray}
This integral can then be calculated within the KPM for all impurity indices $i,j$ and orbital indices $\alpha,\beta$. To give the non-expert the basic idea of the KPM we briefly summarize the concept. The Hamiltonian $\hat{H}$ is approximated in an infinite series of Chebyshev polynomials $T_{n}(\hat{H})$ of order $n$ defined on the interval $[-1,1]$ and thus the spectrum of $\hat{H}$ has to be linearly remapped. In practice this series must be truncated at some point and the respective recursion relation (not shown) is evaluated for the moments $\mu_{n}$ (expansion coefficents) up to a certain order. By using a cosine discretization for the energy-axis, the expanded function can be reconstructed by a fast Fourier transform of the $\mu_{n}$ and in order to prevent so called Gibbs-oscillations of the expanded functions it is necessary to include a multiplicative kernel, too. We use the Jackson kernel as suggested in Ref.\cite{RevModPhys.78.275} resulting in a Gaussian representation of peaks in the spectral density. The broadening $\delta^{n}(\omega)$ of these peaks is analytically connected to the order of expansion and in addition energy dependent, which results in an unphysical inhomogeneous broadening not connected to any quasiparticle lifetimes. A reasonable choice for the correct order $n$ of expansion might be given by the condition, that the average broadening $\langle\delta^{n}(\omega)\rangle$ is smaller than half the energy difference $\delta_{E}$ of the spectrum of $\hat{H}$. Of course one can choose $n$ less or higher depending on the purpose, e.g. producing a smooth density of states or resolving every spectral peak by itself.

In our calculations for the Heisenberg exchange integrals we have checked (not shown) the needed order of expansion for a bulk GaAs system with $N_{x}=N_{y}=N_{z}=N=16$ unit cells in each spatial dimension and decided to use approximatively half the order recommended because the numerical calculations are still very demanding on a cluster using a highly parallelized implementation. In disordered systems we would expect an even higher number than recommended. For calculations concerning the density of states (DOS) we chose the number of moments $\mu_{n}$ so that a smooth function results.

\section{Results}
\label{sec:2}

\subsection{Electronic properties}
\label{sec:2a}

\begin{figure*}
\centering
\resizebox{0.75\textwidth}{!}{\includegraphics{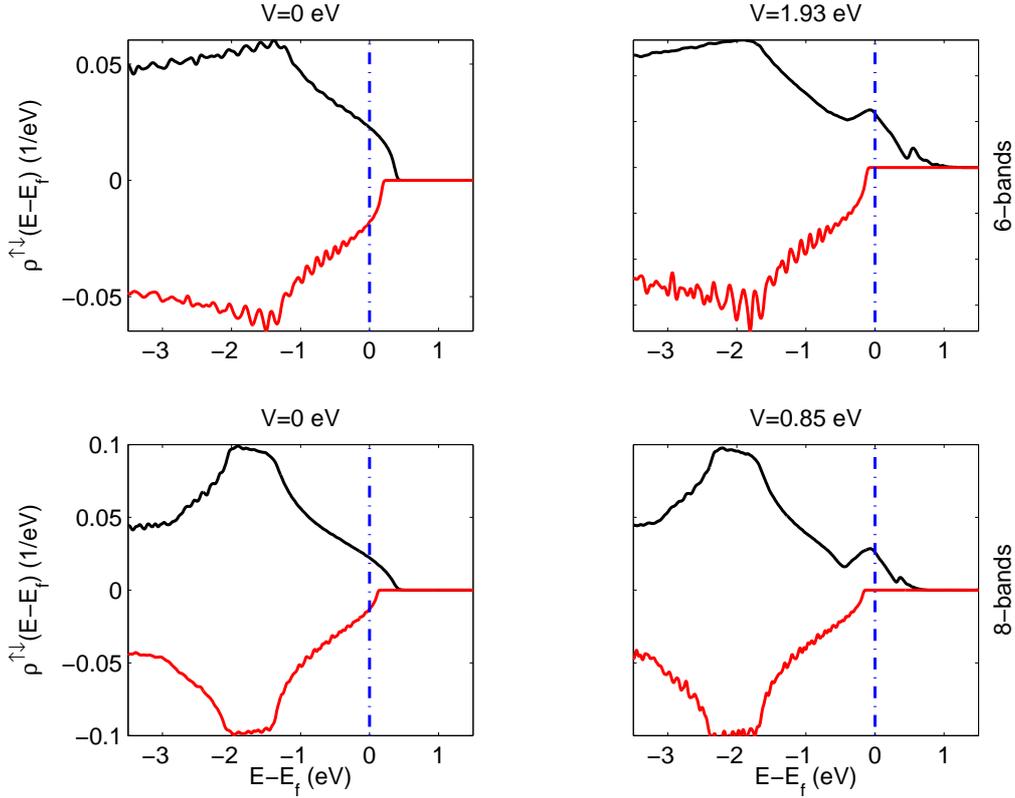}}
\caption{Disorder averaged density of states (DOS) of the \textit{valence bands} calculated for Ga$_{1-x}$Mn$_x$As with $x$=5\% within different tight-binding models with and without the nonmagnetic on-site scattering term $V$. The solid lines correspond to the total DOS while the spin-up (black) / down (red) sector is plotted in the upper and lower part respectively. The system size is N=40 (256 000 lattice sites) and the position of the Fermi-level is denoted with the blue dashed-dotted vertical line.}
\label{fig:2}
\end{figure*}

First we present results for the disorder averaged, spin-resolved total density of states (DOS) of the valence bands (VBs) for Ga$_{1-x}$Mn$_x$As with $x$=5\% with and without the nonmagnetic on-site scattering term $V$ in Fig.\ref{fig:2}. The magnitude of $V$ is chosen to correctly reproduce the acceptor level in the gap for Mn$^{2+}$ of $E_b\approx113$ meV, Ref.\cite{PhysRevB.55.6938}, for one single impurity in the GaAs matrix which is achieved by setting $V=1.93$ eV for the 6-band and $V=0.85$ eV for the 8-band ETBM. The numerical calculations were performed for a system of $N=40$ conventional unit cells in each direction (i.e. 4$\cdot$N$^3$=256 000 lattice sites altogether because of 4 sites per conventional unit cell in the fcc lattice) with 2048 moments and the disorder average was stopped after the fluctuations were less than 1\%.

Let us start by discussing the 6-band case with $V=0$ eV; one can observe, that there is no preformed impurity band (IB) present but a nonvanishing DOS for $E-E_{f}>0$ in combination with a small spin-splitting. Furthermore, the Fermi-level $E_f$ lies within the valence band. In contrast, choosing a finite $V=1.93$ eV changes the situation completely. Now an impurity band formation can clearly be observed around the acceptor level and for $x=5\%$ it is merged with the VB. Due to the fact that we do not apply any approximations in terms of disorder, there is a pronounced sidepeak present which can be attributed to the formation of impurity clusters. Please note, that these peaks vanish if one uses the coherent-potential approximation for disorder. The Fermi-level lies within the IB near the acceptor level, but in the spin-down bandgap, and the spin-splitting is largely enhanced. From these facts one can conclude, that the system could show an insulating character if the eigenstates in the impurity band are localized. In the 8-band case the situation is qualitatively similar, but as we include more bands, the structure and the bandwidth of the DOS is much more realistical.

\begin{figure*}
\centering
\resizebox{0.75\textwidth}{!}{\includegraphics{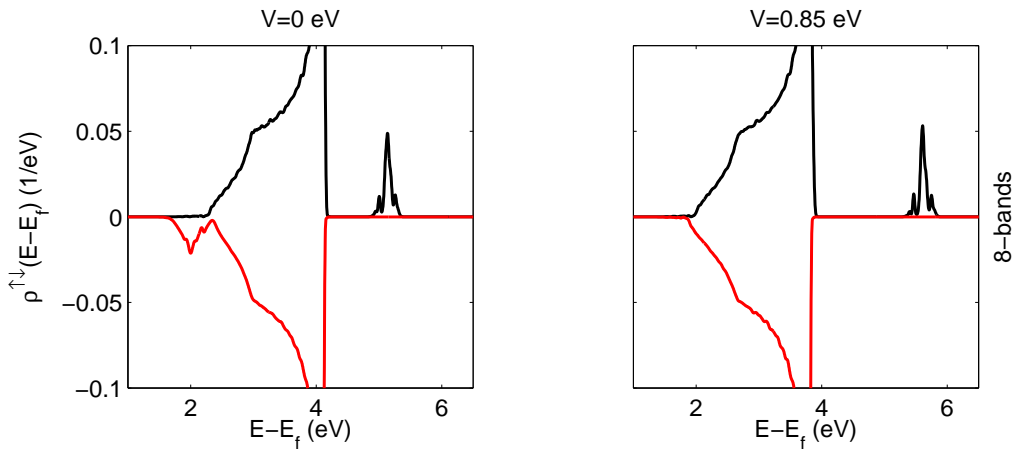}}
\caption{Disorder averaged density of states (DOS) of the \textit{conduction band} calculated for Ga$_{1-x}$Mn$_x$As with $x$=5\% within the 8-band model with and without the nonmagnetic on-site scattering term $V$. The solid lines correspond to the total DOS while the spin-up (black) / down (red) sector is plotted in the upper and lower part respectively. The system size is N=40 conventional unit cells in each direction (256 000 lattice sites).}
\label{fig:3}
\end{figure*}

So far we have analyzed the VBs only and as a next step the conduction band is adressed and visualized in Fig.(\ref{fig:3}) for the 8-band model. Here one observes that in the case of $V=0$ eV the spin-down DOS is enhanced around about $E-E_{f}\approx2$ eV and for $V=0.85$ eV this effect does not occur. In both cases there is also a pronounced impurity band with sidepeaks in the spin-up sector. This is a consequence of using $J_{\alpha}=J_{pd}$ and $V_{\alpha}=V$ for all orbital indices $\alpha$, also for the conduction band. In the case of $V=0.85$ eV the CB impurity band is slightly shifted by $\approx0.5$ eV to higher energies as can be seen from Fig.\ref{fig:2}. Thus the former spin-down DOS enhancement for $V=0$ eV vanishes in the case of $V=0.85$ eV and merges into the GaAs CB.

\subsection{Effective exchange integrals}
\label{sec:2b}
\begin{figure*}
\centering
\resizebox{0.6\textwidth}{!}{\includegraphics{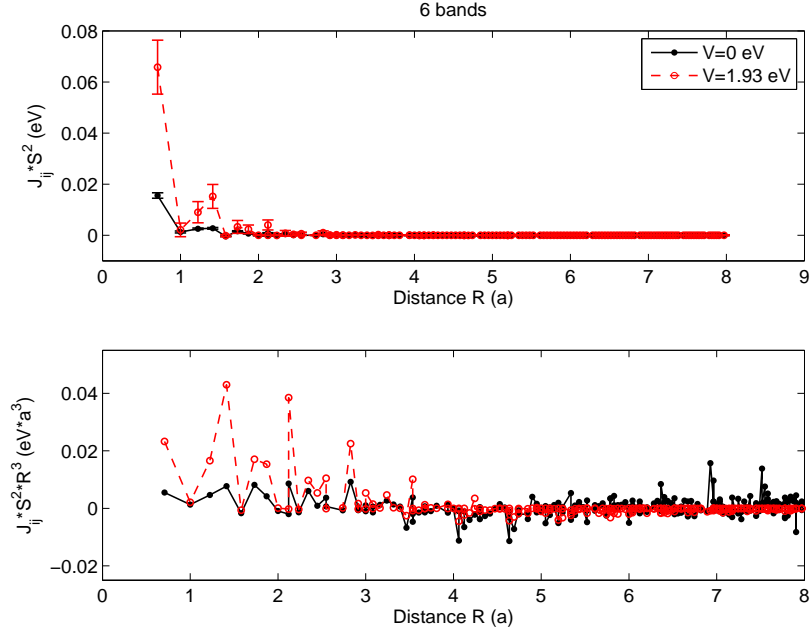}}
\caption{Disorder averaged effective Mn-Mn exchange integrals $J_{ij}$ calculated for Ga$_{1-x}$Mn$_x$As with $x$=5\% within the \textit{6-band model} with and without the nonmagnetic on-site scattering term $V$. The calculations were performed for a system size of N=16 (16384 lattice sites) up to 245 shells with 4096 moments.  In the lower picture $J_{ij}$ was rescaled with the RKKY factor $R^3$ and the error bars are not plotted in order to visualize the long-range behaviour more clearly.}
\label{fig:4}
\end{figure*}

\begin{figure*}
\centering
\resizebox{0.6\textwidth}{!}{\includegraphics{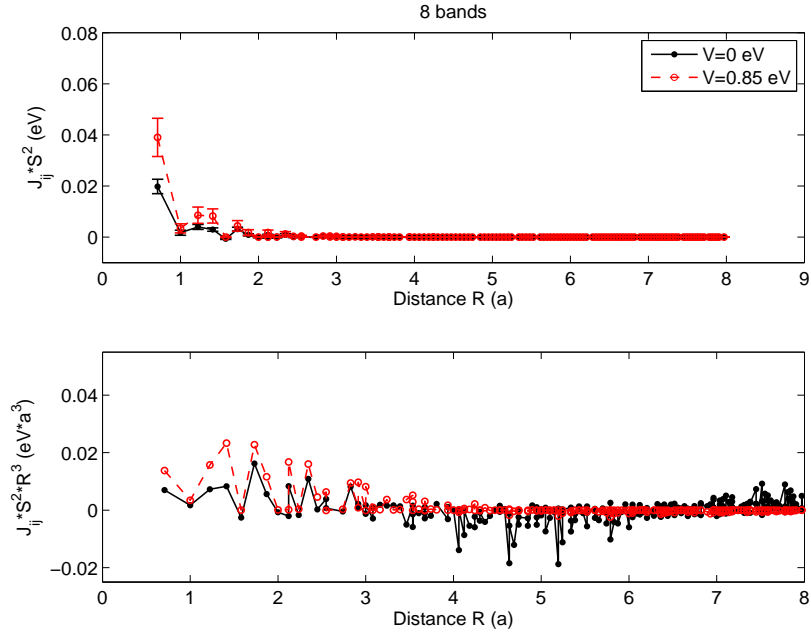}}
\caption{Disorder averaged effective Mn-Mn exchange integrals $J_{ij}$ calculated for Ga$_{1-x}$Mn$_x$As with $x$=5\% within the \textit{8-band model} with and without the nonmagnetic on-site scattering term $V$. The calculations were performed for a system size of N=16 (16384 lattice sites) up to 245 shells with 4096 moments. In the lower picture $J_{ij}$ was rescaled with the RKKY factor $R^3$ and the error bars are not plotted in order to visualize the long-range behaviour more clearly.}
\label{fig:5}
\end{figure*}

The effective exchange integrals are plotted in Fig.\ref{fig:4} for the 6-band model and in Fig.\ref{fig:5} for the 8-band model. The numerical calculations were performed for a system of $N=16$ (16 384 lattice sites) with 4096 moments and the disorder average was carried out over 144 different configurations of disorder while each data point is an average over 1008 values. Please note, these calculations are seriously more demanding than calculating a DOS in terms of computing time and are reasonably performed on a cluster using a parallel implementation. We remark, that the positional disorder average must be carried out over equivalent difference vectors $\mathbf{R}_{ij}=\mathbf{r}_j-\mathbf{r}_i$ with respect to the symmetry of the underlying lattice. For instance, an incorrect average over the absolute distance $R=\vert\vert\mathbf{R}_{ij}\vert\vert$ would correspond to an incorrect spherical symmetry instead of the correct fcc-lattice structure. Furthermore we close the system with periodic boundary conditions and thus the correct average has to take this into account as well. So we use the shortest distance between two Mn-impurities at $\mathbf{r}_i$ and $\mathbf{r}_j$ and difference vectors whose components are larger in magnitude than half the numerical box-size $N/2$ are periodically remapped. Also, we calculate the Heisenberg exchange integrals only up to this absolute magnitude of $R=\vert\vert\mathbf{R}_{ij}\vert\vert\leq N/2$ because equivalent difference vectors with $R>N/2$ can never be realized for a given box size $N$ and the average is then affected too and considered as not reliable.

Comparing the results obtained for the 6-band model one can observe that with and without the potential scattering term the exchange integrals are quite different.
For $V=1.93$ eV the couplings are mainly ferromagnetic and at distances $R\le4a$ their magnitude is enhanced  while at $R>4a$ a damping seems to occur as can be identified in the lower plot in Fig.\ref{fig:3} where the couplings have been rescaled by the RKKY factor $R^3$. Contrary, if the on-site scattering term is neglected, the effective exchange integrals are in general lower in magnitude and the envelope part seems to oscillate around zero. The long-range nature is only visible in the rescaled exchange integrals. Coming back to the former discussion, the results for the couplings $J_{ij}$ agree well with the previous electronic picture given. As the Fermi-level lies in the VB for $V=0$ eV and it is known from textbook RKKY theory for metals, that one should expect a long-range oscillating behavior of the effective exchange integrals in this case, our results are not surprising. If the Fermi level falls into the valence band in a region with delocalized states, RKKY like oscillations can be expected. But if the Fermi level falls into an impurity band, in which the eigenstates may have a more localized character, quite a different behavior may result. This directly reflects itself in the damped more short-ranged effective exchange integrals. Now we shall directly compare to the results of the 8-band model in Fig.(\ref{fig:5}) and we see, that the qualitative features are identical. For $V=0.85$ eV the couplings are slightly reduced in magnitude at distances $R\le4a$ in comparison to the 6-band case with finite $V$ due to the increased weight of the impurity DOS at the Fermi-level. In addition, for $V=0$ eV the long-range oscillating tail seems to be more pronounced as the envelope part is of larger magnitude at $R>4a$ compared to the 6-band model. Overall the picture of the exchange integrals is consistent within the 6- and 8-band model.

\begin{figure*}
\centering
\resizebox{0.6\textwidth}{!}{\includegraphics{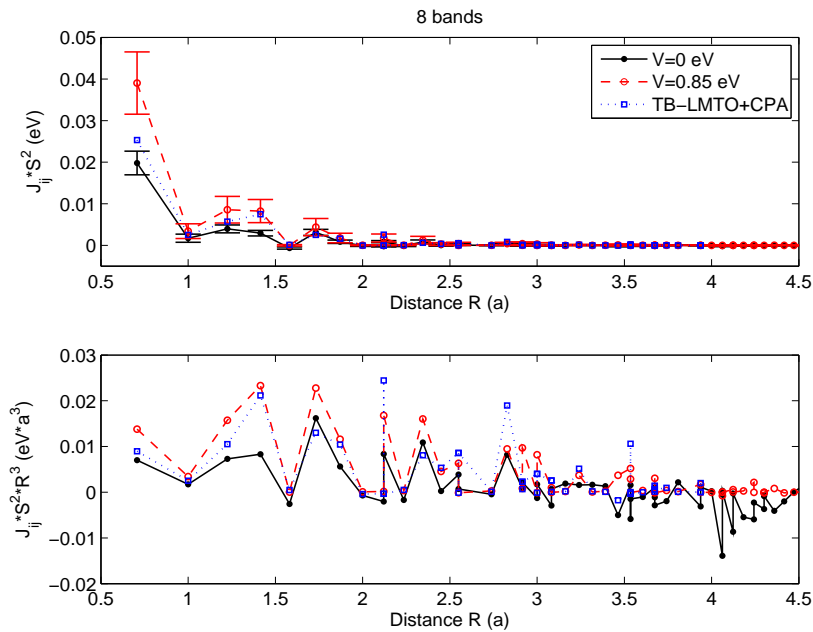}}
\caption{Disorder averaged effective Mn-Mn exchange integrals $J_{ij}$ calculated for Ga$_{1-x}$Mn$_x$As with $x$=5\% within the \textit{8-band model} with and without the nonmagnetic on-site scattering term $V$ in comparison to ab-inito results taken from Ref.\cite{PhysRevB.69.115208}.}
\label{fig:6}
\end{figure*}

At last we want to compare our results to the exchange integrals obtained by ab-initio approaches, see Ref.\cite{PhysRevB.69.115208} for more details, in Fig.(\ref{fig:6}). Here a direct comparison between our 8-band model to results obtained within the TB-LMTO+CPA theory is made. One observes, that the overall shape and structure of the exchange integrals is very similiar for distances up to $\approx3.25 a$ and only the magnitude is different as can be seen from the lower plot in Fig.(\ref{fig:6}). In the case of $V=0$ eV the couplings are mainly lower in value and due to the oscillating nature for larger distances, the discrepancy becomes visible at about $3-3.5 a$.
For $V=0.85$ eV the agreement seems to be better, since the exchange integrals are larger and thus more close to the TB-LMTO+CPA results. In particular the 2nd, 4th, 5th and even higher shells do almost agree in absolute numbers. From the fact that long-range RKKY-like oscillations are absent, the 8-band model with $V=0.85$ eV seems to be a reasonable model to calculate effective Heisenberg exchange integrals and to study magnetic properties. Though the 6-band model with $V=1.93$ eV does also have in principle the same characteristics, it provides short-ranged ferromagnetic couplings which are partially double the magnitude compared to the 8-band approach. Thus from a model point of view, the 6-band V-J model is quantitatively less accurate. But the general agreement between our model calculations and the ab-initio theory is surprisingly good both in a qualitative and a quantitative manner though we applied a lot of approximations on the initial hamiltonian.

\subsection{Critical temperatures}
As mentioned, the $J_{ij}$ calculated and presented in Figs. 4-6 can be used as the parameters of an effective disordered Heisenberg model, for which the Curie temperature $T_C$ can be calculated. The most simple estimate for $T_C$ is obtained from the standard mean-field approximation (MFA) for classical spins as applied in Ref.\cite{PhysRevB.69.115208}.  For our $J_{ij}$ obtained for  the 8-band model and $x=5 \%$  Mn-concentration this estimate yields values  for $T_C^{MF}$ of the magnitude between 160 K for $V=0$ and 390 K for $V \neq 0$, which is even larger than that obtained for the ab-initio $J_{ij}$ from Ref. \cite{PhysRevB.69.115208} ( $\approx$ 290 K for $x = 0.05$). However, it is clear that the MFA overestimates the true $T_C$ and that more sophisticated treatments \cite{RevModPhys.82.1633,Bouzerar_2007_sclrpa,springerlink:10.1140/epjb/e2011-20320-x,PhysRevB.71.113204} of the disordered Heisenberg model have to be applied. From preliminary studies in Ref.\cite{2011arXiv1107.4694B} (using only the 6-band model and a smaller system treated by exact diagonalization) we expect realistic $T_C$-values to be of the magnitude $T_C \le 200$ K.

\section{Summary}
\label{sec:3}
In this work the effective Heisenberg exchange integrals $J_{ij}$ between Mn$^{2+}$ impurities were numerically calculated for  Ga$_{1-x}$Mn$_x$As with $x$=5\% by using two different tight-binding models for the host semiconductor. A comparison on the relevance of a nonmagnetic scattering term $V$ was made with the result, that within the 6-band model for the valence bands as well as in the 8-band model including the conduction band a neglection of $V$ gives rise to a long-range oscillating RKKY-like tail in the couplings $J_{ij}$. In contrast, the inclusion of a finite nonmagnetic on-site scattering term reproduces the Mn-acceptor level in the limit of one impurity and changes the nature of the couplings completely to a short-ranged and damped character and yields mainly positive values. This behavior was explained by analyzing the total spin-resolved density of states, because by choosing $V=0$ eV the Fermi-level lies in the valence band and for $V>0$ eV an impurity band formation around the Mn-acceptor level occurred in combination with the Fermi-level lying therein. In comparison to available ab-initio results, our calculated couplings are of qualitatively of the same structure and do even quantitatively agree. Thus it is important for ferromagnetism to become possible to include the potential scattering term.
%
%
%
%
\bibliographystyle{epj}
\bibliography{epjb_refs}
%
%
%

\end{document}